\documentclass[final,3p]{elsarticle}

\usepackage{graphicx,url}

\usepackage{esint}
\usepackage[T1]{fontenc}
\usepackage[latin9]{inputenc}

\usepackage{amssymb}

\biboptions{sort&compress} 

\newcounter{bla}

\newcommand{\be}{\begin{equation}}
\newcommand{\ee}{\end{equation}}
\newcommand{\ba}{\begin{eqnarray}}
\newcommand{\ea}{\end{eqnarray}}

\def\lsim{\raise0.3ex\hbox{$\;<$\kern-0.75em\raise-1.1ex\hbox{$\sim\;$}}}
\def\gsim{\raise0.3ex\hbox{$\;>$\kern-0.75em\raise-1.1ex\hbox{$\sim\;$}}}

\def\theta{\vartheta}

\journal{Computer Physics Communications}

\begin{document}

\begin{frontmatter}



  \title{{\tt AAfrag}: Interpolation routines for Monte Carlo results on
    secondary production in proton-proton, proton-nucleus and nucleus-nucleus
    interactions}


\author[a]{M.~Kachelrie\ss}
\author[b]{I.~V.~Moskalenko}
\author[c,d]{S.~Ostapchenko}
\address[a]{Institutt for fysikk, NTNU, Trondheim, Norway}
\address[b]{Hansen Experimental Physics Laboratory \& 
Kavli Institute for Particle Astrophysics and Cosmology, 
Stanford University, Stanford, CA 94305, U.S.A}
\address[c]{Frankfurt Institute for Advanced Studies, Frankfurt, Germany}
\address[d]{D.~V.~Skobeltsyn Institute of Nuclear Physics,
 Moscow State University, Russia}

\begin{abstract}
We provide a compilation of predictions of the QGSJET-II-04m model for the production of secondary species (photons, neutrinos, electrons, positrons, and antinucleons) that are covering a wide range of energies of the beam particles in proton-proton, proton-nucleus, nucleus-proton, and nucleus-nucleus reactions. 
The current version of QGSJET-II-04m has an improved treatment of the production of secondary particles
at low energies: the parameters of the hadronization procedure have been fine-tuned, based on a number of recent benchmark experimental data, notably, from the LHCf, LHCb, and NA61 experiments. Our results for
the production spectra are made publicly accessible through the interpolation routines {\tt AAfrag} which are described below. Besides, we comment on the impact of Feynman scaling 
violation and isospin symmetry effects on antinucleon production.
\end{abstract}


\end{frontmatter}



{\bf PROGRAM SUMMARY}

\begin{small}
\noindent
{\em Manuscript Title:}
    {\tt AAfrag}: Interpolation routines for Monte Carlo results of
    secondary production in proton-proton, proton-nucleus and nucleus-nucleus
    interactions\\
 {\em Program Title:} {\tt AAfrag 1.01} 
\\
{\em Journal Reference:}                                      \\
{\em Catalogue identifier:}                                   \\
{\em Licensing provisions:}                                   
CC by NC 3.0.  
\\
{\em Programming language:}  Fortran 90                      \\
{\em Computer:}                                               
Any computer with Fortran 90 compiler                  \\ 
{\em Operating system:}  Any system with Fortran 90 compiler                \\
{\em RAM:} 72 Mbytes                                              \\
{\em Supplementary material:}                                
see \url{http://aafrag.sourceforge.io}
{\em Classification:}
1.2 	Nuclear Processes, 11.2 Phase Space and Event Simulation\\
{\em Nature of problem:}
Calculation of secondaries (photons, neutrinos, electrons, positrons,
 protons, and antiprotons) produced in hadronic interactions
   \\
{\em Solution method:}
Results from the Monte Carlo simulation QGSJET-II-04m are interpolated.
\end{small}

\section{Introduction}

A large variety of applications in astrophysics and astroparticle physics relies on a precise knowledge of the cross sections for production of secondary species in hadronic interactions. 
Prominent examples are indirect
dark matter searches using 
cosmic rays (CRs), where a putative signal represents an excess
over the regular background 
produced in
astrophysical processes~\cite{IDM}, and
the spectrum of CRs deciphered from observations of the electromagnetic emission produced in their interactions in the interstellar gas~\cite{CRspec}. 
Such applications typically employ empirical parameterizations of secondary production cross sections tuned to
available accelerator data~\cite{tan1983,add}. Despite
the evident convenience of such parameterizations, this approach
has a number of caveats. In particular, it usually relies on empirical
scaling laws when extrapolating outside the measured kinematic range. Meanwhile,
high-energy extrapolations of such empirical parameterizations are generally
unreliable. Besides, this approach does not work well when there is a variety of nuclei species in CRs (beam) and in the interstellar gas (target). The latter is usually cured by an application of the so-called
empirical ``nuclear enhancement'' factor, which varies substantially between
different publications~\cite{K}. Meanwhile, as it has been demonstrated in
Ref.~\cite{kmo2014}, properly defined nuclear enhancement factors depend
both on the nature of interacting particles and on the spectral slopes of the
CR spectra of individual nuclei.

A number of serious issues associated with the traditional approach prompted us
to reevaluate the production of secondaries using Monte Carlo generators for
hadronic interactions, that describe the production of secondary photons, neutrinos, positrons, antiproton and antineutrons in proton-proton, proton-nucleus, and nucleus-nucleus
collisions consistently within the same framework~\cite{kmo2014,ko2012,ko2014,kmo2015}. To this
end, a special effort has been made to improve the low-energy
behavior of the QGSJET-II-04  Monte Carlo generator~\cite{ost11},
which is widely used in CR astrophysics to treat high-energy interactions. 
In particular, we aimed at the accurate description of antiproton
production~\cite{kmo2015} that is of special interest for indirect dark matter searches.

In this work, we provide a compilation of predictions of the QGSJET-II-04m model \cite{kmo2015} for the production of secondary species (photons, neutrinos, electrons, positrons, antiprotons and antineutrons) that are covering a wide range of energies of the beam particles in proton-proton, proton-nucleus, nucleus-proton, and nucleus-nucleus reactions. 
Compared to Ref.~\cite{kmo2015}, the current version of QGSJET-II-04m has an improved treatment of the production of secondary particles at low energies: the parameters of the hadronization procedure have been fine-tuned, based on a number of recent benchmark experimental data, notably, from the LHCf, LHCb, and NA61 experiments. Our results for the production spectra are made publicly available through the interpolation routines {\tt AAfrag} which are described below. 

\begin{figure}
\centering{}
\includegraphics[scale=0.85]{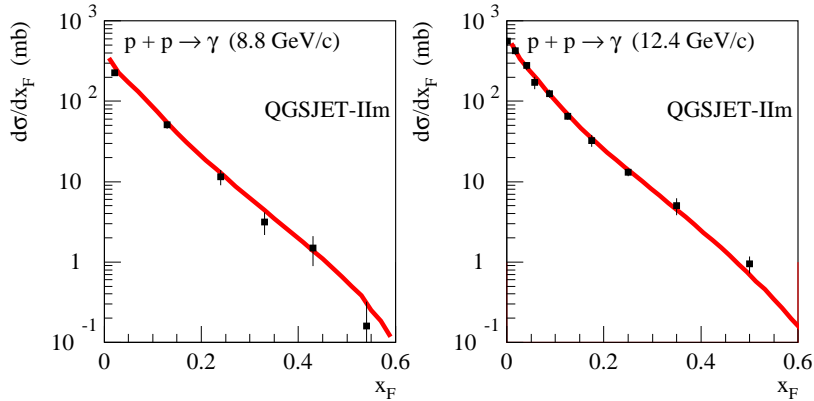}
\caption{Calculated Feynman $x$ spectra 
 $d\sigma_{pp}^{\gamma}/dx_{{\rm F}}$ of photons (red line)
in the c.m.\ frame for $pp$-collisions at 8.8 GeV/$c$ (left) and 12.4 GeV/$c$
(right), compared to experimental data (black errorbars) from 
Refs.~\cite{boo1983,jae1973}.\label{fig:pp8}}
\end{figure}

\section{Selected results}

\begin{figure}
\centering{}
\includegraphics[scale=0.7]{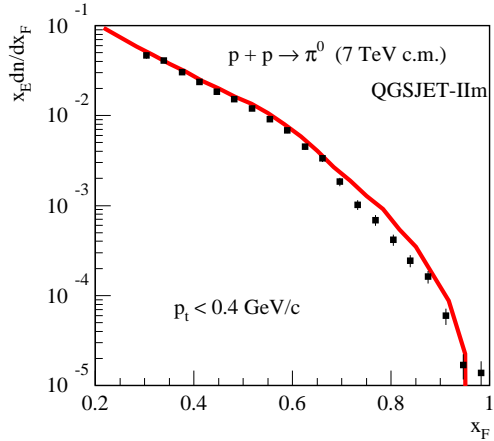}
\caption{Calculated invariant energy spectrum
$x_{E}\, dn_{pp}^{\pi^{0}}/dx_{{\rm F}}$,  with $x_{E}=2E/\sqrt{s}$, 
of neutral pions (red line) in the c.m.~frame for $pp$-collisions at
$\sqrt{s}=7$ TeV, compared to the data (black errorbars) of the LHCf
experiment~\cite{lhcf-pi0}.
\label{fig:lhcf}}
\end{figure}

\begin{figure}[t]
\centering{}
\includegraphics[scale=0.8]{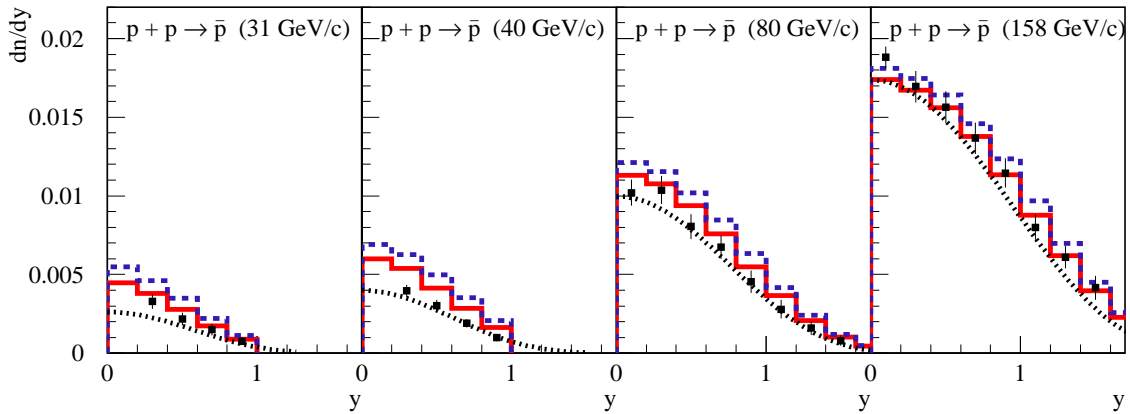}
\caption{Calculated rapidity distributions $dn/dy$ of antiprotons (solid red
  lines) and antineutrons (dashed blue lines) for $pp$-collisions at different
incident momenta (from left to right: 31, 40, 80, and 158 GeV/$c$), compared
to  NA61 data on $\bar p$-production~\cite{na61} (black errorbars). 
Additionally the results for the parameterization~\cite{tan1983} are shown
as dotted black lines.
\label{fig:pp-ap}}
\end{figure}

\begin{figure}
\centering{}
\includegraphics[scale=0.8]{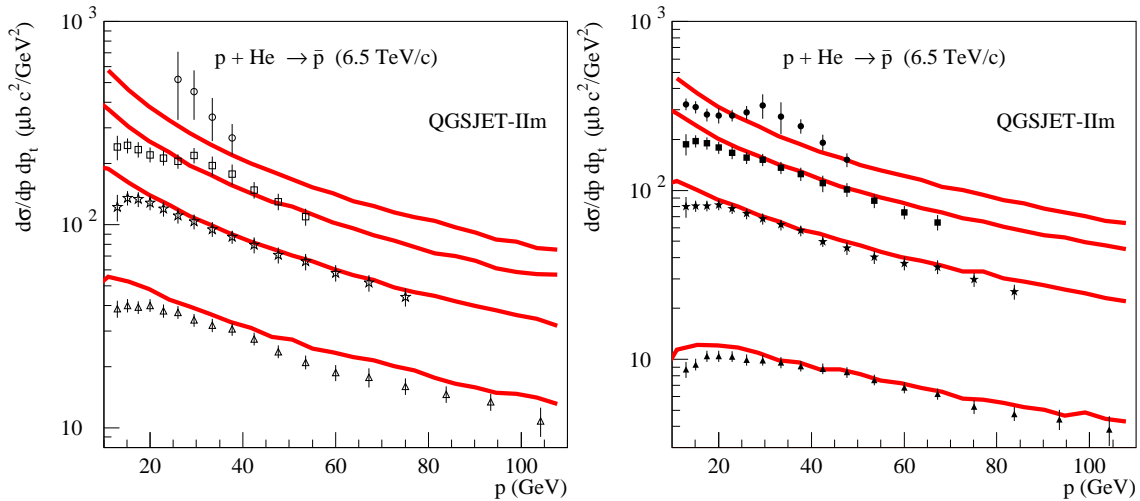}
\caption{Calculated laboratory momentum spectra of antiprotons, 
$d\sigma_{p{\rm He}}^{\bar{p}}/dp\, dp_{{\rm t}}$,
for $p{\rm He}$-collisions at 6.5 TeV/c for different $p_{{\rm t}}$ intervals 
[from top to bottom:
 0.4--0.6,  0.7--0.8, 0.9-1.1,  1.2--1.5\,GeV (left) and
  0.6--0.7,  0.8--0.9,  1.1--1.2,  1.5--2.0\,GeV (right)],
 compared to  data of the LHCb experiment~\cite{aai2018}.
\label{fig:lhcb}}
\end{figure}

The quality of the description of secondary particle production
by the QGSJET-II-04m model is illustrated in Figs.~\ref{fig:pp8}--\ref{fig:lhcb}.
In Fig.~\ref{fig:pp8}, the results of 
this model for the spectra of photons as function of the
Feynman scaling variable $x_{\rm F}=2p_\|/\sqrt{s}$ in the center-of-mass
(c.m.)~frame
are compared to the corresponding experimental data for $pp$-collisions
at rather low incident proton momenta, 8.8 and
12.4\,GeV/c~\cite{boo1983,jae1973}. In turn, in Fig.~\ref{fig:lhcf},
we plot the calculated spectrum of neutral pions in the c.m.~frame
together with the data of the LHCf experiment~\cite{lhcf-pi0}
for proton-proton collisions at 7\,TeV c.m.~energy.
Both the low- and high-energy data sets are well described by the results
of the QGSJET-II-04m model.

Concerning low energy antiproton ($\bar p$) production, important constraints
on hadronic interaction models are provided by the recent results of the NA61
experiment~\cite{na61}. The calculated rapidity distributions of the
produced antiprotons in the c.m.\ frame (solid red lines) are compared in
Fig.~\ref{fig:pp-ap} to the experimental data (black errorbars)
for $pp$-collisions at different incident proton momenta: 31, 40, 80, and
158 GeV/$c$. Additionally the results for the parameterization~\cite{tan1983}
are shown as dotted black lines.
As discussed previously in Ref.~\cite{kmo2015}, a significant uncertainty
for the predicted flux of cosmic ray antiprotons at low ($E_{\bar{p}}^{{\rm kin}}<10$\,GeV) energies is caused by the rather weak experimental constraints
on the $\bar p$ yield from proton-proton collisions close to the production
threshold. In particular, the substantially larger  $\bar p$ yield predicted
by the  QGSJET-II-04m model in that energy range gave rise to an almost
factor two higher antiproton flux at $E_{\bar{p}}^{{\rm kin}}=1$\,GeV~\cite{kmo2015}, compared to the calculations based on the parameterization
of Ref.~\cite{tan1983}. This uncertainty is now greatly reduced by the NA61 data, e.g., as one can see in the leftmost panel of Fig.~\ref{fig:pp-ap}:
The strong suppression of the $\bar p$ yield near the threshold,  which is
predicted by the parameterization of Ref.~\cite{tan1983}, is not supported by
the experimental data.

A very important high energy benchmark are studies of proton-helium collisions 
by the LHCb experiment, performed in the fixed target mode~\cite{aai2018}.
A comparison of our calculations with the corresponding experimental
data for the momentum distribution of secondary antiprotons in the laboratory
frame for proton-helium collisions at 6.5 TeV/c is plotted in 
Fig.~\ref{fig:lhcb}. There is a generally satisfactory agreement with the
data, particularly at relatively large antiproton momenta
which is of main importance for astrophysical applications.
While previously available experimental data on very high energy  $\bar p$ 
production were restricted to central rapidity range, the LHCb results provide
an important baseline for high energy extrapolations of forward production
spectra of antiprotons.

\section{Feynman scaling violation and isospin symmetry}

As discussed in Ref.~\cite{kmo2014,kmo2015}, the differences between various
interaction models or parameterizations of production spectra in the
predicted secondary CR fluxes can be conveniently analyzed using
the so-called $Z$-factors. These are defined as the spectrally
averaged energy fraction transferred to the produced particles of a given
type in proton-proton, proton-nucleus, or nucleus-nucleus interactions.
$Z$-factors may be introduced when the spectra of ambient CRs in
the relevant energy range can be approximated by a power law. For the
production of secondary CRs of the type $X$
($X=\gamma,e^{\pm},\nu,\bar{p},\bar{n})$
in collisions of a primary particle $i$ with the gas component $j$ of the
interstellar medium, the $Z$-factor is defined as~\cite{gaisser}
\be
Z_{ij}^{X}(E_{X},\alpha_{i})=\int_{0}^{1}\! dz\; 
z^{\alpha_{i}-1}\,\frac{d\sigma_{ij}^{X}(E_{X}/z,z)}{dz}\,,
\ee
where $\alpha_{i}$ is the slope of the primary CR spectrum, $z$ denotes the
energy fraction transferred to the particles $X$, and 
$d\sigma_{ij}^{X}(E_0,z)/dz$ the corresponding production spectrum.
For the assumed case of a  power-law primary flux, the partial contribution
to the secondary flux of the particles $X$ factorizes into a product of the
primary flux  $I_{i}$ and the $Z$-factor,
\be
q_{ij}^{X}(E_{X})\propto I_{i}(E_{X})\, Z_{ij}^{X}(E_{X},\alpha_{i})\,.
\ee
Thus all the dependence on the interaction model in use is contained
in the $Z$-factor $Z_{ij}^{X}$.

\begin{figure}
\centering{}
\includegraphics[scale=0.8]{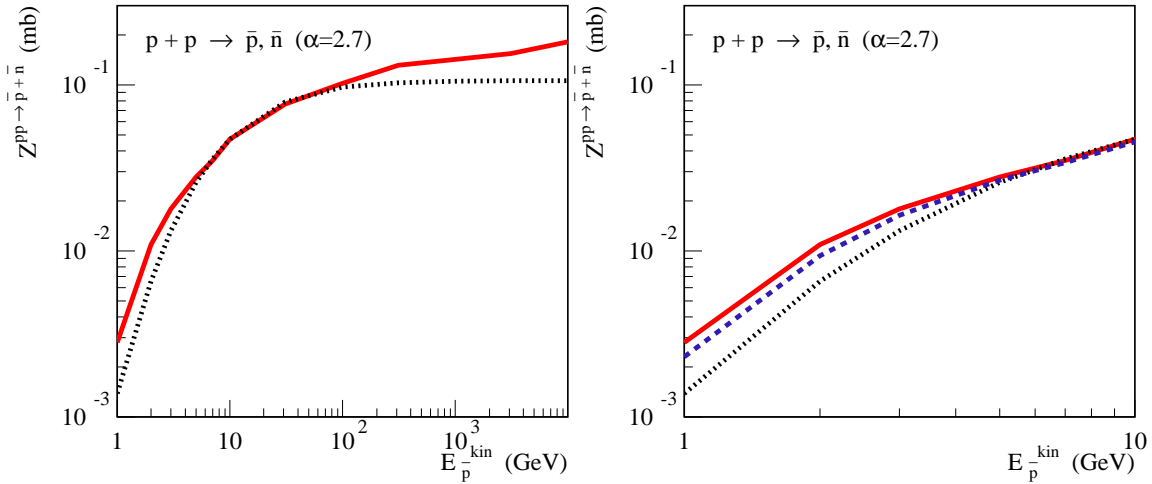}
\caption{Left: $Z_{pp}^{\bar{p}+\bar{n}}$  calculated using the QGSJET-II-04m
model (red solid line) and $2Z_{pp}^{\bar{p}}$
calculated using the parameterization of Ref.~\cite{tan1983} (black  dotted
line) -- both for the  primary proton spectral slope $\alpha=2.7$.
The right panel zooms into the low-energy part and contains also
 $2Z_{pp}^{\bar{p}}$ calculated using  QGSJET-II-04m (blue dashed line).
\label{fig:zfac}}
\end{figure}

\begin{figure}
\centering{}
\includegraphics[scale=0.8]{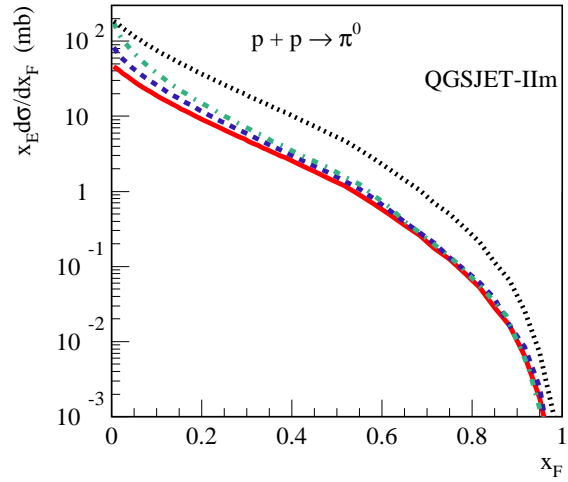}
\caption{Invariant energy spectrum of neutral pions,
 $x_{E}\, d\sigma_{pp}^{\pi^{0}}/dx_{{\rm F}}$,
in c.m.~frame for $pp$-collisions at $\sqrt{s}=10^{2}$, $10^{3}$,
and $10^{4}$ GeV -- solid, dashed, and dashed-dotted lines respectively.
Dotted line corresponds to a rescaling of the spectrum from $\sqrt{s}=10^{2}$
GeV to $\sqrt{s}=10^{4}$ GeV, according to the rise of the height
of the central rapidity plateau.\label{fig:pi0-lim}}
\end{figure}

For illustration, in Fig.~\ref{fig:zfac} we compare  our results for the
energy dependence of $Z_{pp}^{\bar{p}+\bar{n}}$ for $\alpha=2.7$,
shown by the red solid line, with the one obtained from the parameterization
of $\bar{p}$ production from Ref.~\cite{tan1983}, shown as a black dotted line.
In the latter case, equal production yields of $\bar{p}$ and $\bar{n}$ are
assumed, i.e.\ the dotted line corresponds to
$2Z_{pp}^{\bar{p}}$. As one can see on the left panel of
Fig.~\ref{fig:zfac}, the main difference between
the two results originates from the violation of the Feynman scaling
in the high energy limit, which is properly accounted for by the QGSJET-II-04m
model. In contrast, the parameterization of Ref.~\cite{tan1983} gives rise
to perfect Feynman scaling at high energies.
It is worth stressing that the origin of these scaling violations
is rather nontrivial, resulting from both the energy-rise of the inelastic
cross section and a modification of the shape of the momentum spectra.
Some recent parameterizations of secondary particle production assume
that the momentum spectra scale at sufficiently high energies (see, e.g.,
Ref.~\cite{blu2017}), i.e.\ that the factorization
\begin{equation}
\frac{x_{E}\, d\sigma_{pp}^{X}(s,x_{{\rm F}})}{dx_{{\rm F}}}
\simeq \xi_{X}(s)\, f(x_{{\rm F}})
\label{eq:scaling}
\end{equation}
holds.
Moreover, they parameterize the resulting $s$-dependence of $\xi_{X}(s)$ using
experimental data on the height of the rapidity plateau in the c.m.~frame,
setting
\begin{equation}
\xi_{X}(s)\propto \left. \frac{d\sigma_{pp}^{X}}{dy}\right|_{y=0}=
\left.x_{E}\, \frac{d\sigma_{pp}^{X}}{dx_{{\rm F}}}\right|_{x_{{\rm F}}=0} .
\end{equation}
Such an assumption leads to a violation of energy conservation in
the high energy limit and gives rise to a considerable overestimation
of secondary CR spectra. This  holds  especially for photons, positrons,
and neutrinos, because an important contribution to their production spectra
comes from the hadronization of  valence quarks of the incident proton.
Because of the increase of multiple scattering~\cite{den2011}, the height of
the central rapidity plateau rises asymptotically as a power of energy,
$\left.d\sigma_{pp}^{X}/dy\right|_{y=0}\propto s^{\Delta_{{\rm eff}}}$,
and thus the factor $\xi_{X}$ in Eq.~(\ref{eq:scaling}) rises with the
same slope. This would imply, in turn, that the average energy fraction
transferred to secondary particles of any given type
$X$, $\langle z_{E}^{X}(s)\rangle\propto\xi_{X}(s)/\sigma_{pp}^{{\rm inel}}(s)$,
has an unlimited rise---which is clearly nonphysical.
In reality, the steep energy rise of the central rapidity
plateau is accompanied by a moderate ``softening'' of the very forward
part of the production spectra, because multiple scattering processes have to
share the energy~\cite{obp2016}, as illustrated in Fig.~\ref{fig:pi0-lim}
for the particular case of neutral pion production.

\begin{figure}
\centering{}
\includegraphics[scale=0.4]{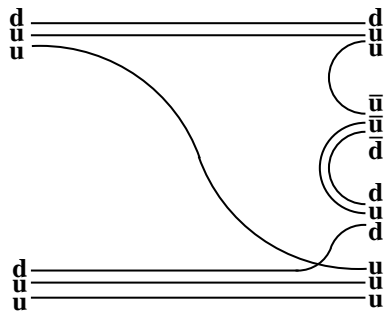}
\caption{Example diagram of the undeveloped cylinder type, which contributes
to antiproton production in $pp$-collisions.\label{fig:uncyl}}
\end{figure}

Let us now discuss the difference between the solid and dotted lines at
low-energies $E_{\bar{p}}^{{\rm kin}}<10$\,GeV which can be seen in the right
panel of Fig.~\ref{fig:zfac}. These differences are partly explained
by the larger $\bar{n}$  (compared to $\bar{p}$) yield  close to the production
threshold. This is illustrated by the blue dashed line in the same panel, which
corresponds to $2Z_{pp}^{\bar{p}}$ from the QGSJET-II-04m model.
At the first sight, such a $\bar{n}/\bar{p}$ enhancement seems surprising:
while isospin symmetry is not an exact one for strong interactions, it
holds with a very good accuracy, thanks to the small mass difference between
the $u$ and $d$ quarks. Taking into account that both $\bar{p}$ and
$\bar{n}$ originate from a central production mechanism, without involving
valence quarks of the colliding protons, one expects equal average
yields of antiprotons and antineutrons.

The origin of the unexpected neutron enhancement is related to the valence
quark content of the proton ($uud$) and the suppression of heavy baryon
production close to the production threshold.
To explain this in more detail, let us remark that in the
low-energy regime particle production in $pp$-collisions is dominated
by the so-called ``undeveloped cylinder'' diagrams, as exemplified
in Fig.~\ref{fig:uncyl}.
In such a case, a valence quark from one of the colliding
protons (projectile $u$-quark in Fig.~\ref{fig:uncyl}) combines
with a diquark from the other proton (target $uu$-diquark in
Fig.~\ref{fig:uncyl})
to form a final baryon. At the same time, a string of color field
is being formed between the remaining valence diquark and quark ($ud-d$
string in Fig.~\ref{fig:uncyl}). With those diquark and quark sitting
on the string ends and flying apart, the string breaks and gives rise
to additional secondary hadron production via a creation of quark-antiquark
and diquark-antidiquark pairs from the vacuum. 

Returning back to our case
of interest, near-threshold production of antiprotons and antineutrons,
we can neglect final states containing heavy baryon resonances and
consider only the $ud\bar{u}\bar{d}$-diquark contribution to this
hadronization process. Using simple quark counting
rules, we are left with only four possible final states with the following
relative weights: 1) $ppp\bar{p}$ -- 2/9; 2) $ppn\bar{n}$ -- 7/18;
3) $pn\bar{p}\Delta^{+\!+}$ -- 5/18; 4) $nn\bar{n}\Delta^{+\!+}$ -- 1/9. 
Accounting for the decay of the $\Delta^{+\!+}$,
this seems to yield equal multiplicities for $\bar{p}$ and $\bar{n}$.
However, according to our reasoning, close to the threshold we have
to neglect the 3rd and 4th configurations of final states since they
are suppressed because of the high mass of the $\Delta^{+\!+}$ resonance.
This immediately
gives rise to an almost factor of 2 enhancement of the $\bar{n}/\bar{p}$
yield. 

It is worth stressing that this effect quickly
fades away as the collision energy  increases. This can
be seen in Fig.~\ref{fig:pp-ap}, where we plot the rapidity distributions of produced antineutrons with dashed blue lines: the relative $\bar{n}/\bar{p}$
excess changes from $\simeq20$\% to $\simeq5$\% when the incident
proton momentum increases from 31 GeV/$c$ to 158 GeV/$c$.
The same effect is also clearly seen in the right panel of Fig.~\ref{fig:zfac}: 
the solid and dashed lines coinside at sufficiently high energies. In contrast, in some
recent parameterizations one assumes a substantial violation of the
isospin symmetry by postulating a large {\em energy-independent\/}
excess of antineutrons (see, e.g., Ref.~\cite{dim2014}).

\section{Program structure and example  output}
\subsection{Program structure}
The program consists of two files {\tt AAfrag101.f90} and {\tt user101.f90}. 
The file {\tt AAfrag101.f90} contains the module {\tt spectra},
the main program,  the initialization and the interpolation subroutines.
For standard applications no changes in this file are necessary. 
The file {\tt user101.f90} contains as an example the calculation of
secondary spectra for incident proton energy $E_p=100$\,GeV.

More precisely, the file {\tt AAfrag101.f90} starts with the module
{\tt spectra} containing the definition of internal variables
and the values of some constants. The following {\tt main program} calls
first the initialization subroutine {\tt init} and then the subroutine
{\tt user\_main} contained in the file {\tt user101.f90}. The latter
subroutine has to be adopted by the users for their specific needs.

The initialization is performed by two subroutines, {\tt init\_low}
and {\tt init\_high}. The latter initializes  a grid logarithmic in
primary energy at primary energies $E_p\geq 10$\,GeV, while the former
defines an additional grid linear in primary energy for the reaction
$k=\{1,2,3,4\}$ at lower energies. 
The two subroutines read the data files for the secondary production
cross sections calculated in various QGSJET-II-04m runs which are collected
in the folder {\tt Tables}.

\begin{table}
  \centering{
\begin{tabular}{|l|c|c|c|c|}
  \hline
  function & $q=1$  & $q=2$  & $q=3$  & $q=4$ \\
{\tt spec\_nu(E\_p,E\_s,q,k)} & $\nu_e$  & $\bar\nu_e$ &$\nu_\mu$ & $\bar\nu_\mu$\\
{\tt spec\_elpos(E\_p,E\_s,q,k)} & $e^-$ & $e^+$ & &\\
{\tt spec\_pap(E\_p,E\_s,q,k)}  & p &$\bar p$ & &\\
{\tt spec\_nan(E\_p,E\_s,q,k)} & $n$ & $\bar n$ & &\\
\hline
  \end{tabular}}
\caption{Dependence of the particle type on the variable $q$.}
\end{table}

The functions 
{\tt spec\_gam(E\_p,E\_s,k)},
{\tt spec\_nu\_tot(E\_p,E\_s,q,k)},  
{\tt spec\_elpos(E\_p,E\_s,q,k)},
{\tt spec\_pap(E\_p,E\_s,q,k)},  and
{\tt spec\_nan(E\_p,E\_s,q,k)} 
interpolate the production spectra $E_s^2 d\sigma^k((E_p,E_s)/dE_s$ of the
secondaries $\{\gamma,\nu,e^-,e^+,\nu_i,p,\bar p,n,\bar n\}$ in
proton-proton ($k=1$), proton-helium ($k=2$), helium-proton ($k=3$), 
helium-helium ($k=4$),
carbon-proton ($k=5$),  aluminium-proton ($k=5$), and iron-proton ($k=6$)
reactions. The connection between the variable $q$ and the particle type is
specified in Table~1, while the mass numbers of target and projectile nuclei
are given for the various reactions in Table~2. Alternatively to the function
{\tt spec\_nu\_tot(E\_p,E\_s,q,k)} which provides the spectrum for an
individual neutrino type, the function
{\tt spec\_nu\_tot(E\_p,E\_s,k,spec)} returns the sum of the $\nu_e$,
$\nu_e$, $\bar\nu_e$ and  $\bar\nu_\mu$ spectra.
Note that we keep (anti-) neutrons
stable. Thus antinucleons can be used for studies of antideuteron production.
Moreover, neutron escape from magnetized sources can be employed in CR
studies.

\begin{table}
  \centering{
\begin{tabular}{|c|c|c|c|c|c|c|c|}
\hline
  reaction & 1 & 2 & 3 & 4 & 5 & 6 & 7 \\ \hline
  beam-target & p-p & p-He & He-p & He-He & C-p & Al-p & Fe-p\\\hline
  mass number & 1-1 & 1-4 & 4-1 & 4-4 & 12-1 & 26-1 & 56-1\\
\hline
  \end{tabular}}
\caption{Beam and target particles implemented in {\tt AAfrag}.}
\end{table}

The primary energy $E_p$ denotes the total energy of the primary nucleus
in GeV, $E_s$ the energy in GeV of the produced secondary, and the
differential production cross sections are measured in mbarn.
The minimal interpolated primary energy varies from $E_p=5$\,GeV for
protons to 100\,GeV for iron, while the maximal energy is
$E_p=10^{20}$\,eV.

\subsection{Example output}

The files {\tt spec\_gam}, {\tt spec\_nu},{\tt spec\_nu\_tot},
{\tt spec\_elpos}, {\tt spec\_aprot}, and
{\tt spec\_aneut} contain the production spectra
$E_s^2 d\sigma^k(E_p,E_s)/dE_s$ of photons, the four neutrinos $\nu_e$,
$\nu_e$, $\bar\nu_e$ and  $\bar\nu_\mu$, their sum, electrons and positrons,
antiprotons, and antineutrons for the case of proton-proton collisions at 
$E_p=100$ GeV.

\section{Summary}

We compared the results of an updated version of QGSJET-II-04m to a wide range
of experimental results covering both low- and high-energy data, finding a
satisfying agreement between the measured and the calculated  energy spectra. We
discussed the reason for the apparent enhancement of the $\bar n/\bar p$
yield at low energies and demonstrated that it is explained by a threshold effect.
Our results for the production spectra of photons, neutrinos, electrons,
positrons, and antiprotons in proton-proton, proton-nucleus, nucleus-proton,
and nucleus-nucleus collisions are made publicly available through the
described interpolation routines {\tt AAfrag}.

\section*{Acknowledgements}
\noindent
S.O. acknowledges support from project OS\,481/2-1 of the 
Deutsche Forschungsgemeinschaft. I.V.M. acknowledges partial support from NASA grant NNX17AB48G.



\begin{thebibliography}{99}

\bibitem{IDM}
J.~Silk and M.~Srednicki,
  Phys.\ Rev.\ Lett.\  {\bf 53}, 624 (1984);
  for reviews see e.g.\
T.~A.~Porter, R.~P.~Johnson and P.~W.~Graham,
  Ann.\ Rev.\ Astron.\ Astrophys.\  {\bf 49}, 155 (2011); 
  M.~Cirelli,
  J.\ Phys.\ Conf.\ Ser.\  {\bf 718}, no. 2, 022005 (2016).

  
\bibitem{CRspec}
  M.~Ackermann {\it et al.} [Fermi-LAT Collaboration],
  Astrophys.\ J.\  {\bf 750}, 3 (2012);
R.~Yang, F.~Aharonian, and C.~Evoli,
  Phys.\ Rev.\ D {\bf 93}, no. 12, 123007 (2016);
A.~Neronov, D.~Malyshev, and D.~V.~Semikoz,
  Astron.\ Astrophys.\  {\bf 606}, A22 (2017).


  
\bibitem{tan1983}
L.\ C.\ Tan and L.\  K.\   Ng, J.\ Phys.\ G {\bf 9}, 1289 (1983).

\bibitem{add}
G.D.~Badhwar, S.~A.~Stephens, and R.~L.~Golden, 
Phys.\ Rev.\ {\bf D5}, 820 (1977);
C.~D.~Dermer, Astron.\ Astrophys.\ {\bf 157}, 223 (1986)


\bibitem{K}
  T.~K.~Gaisser and R.~K.~Schaefer,
Astrophys.\ J.\   {\bf 394},  174 (1992);
M.~Mori,
  Astropart.\ Phys.\  {\bf 31}, 341 (2009).


\bibitem{kmo2014}
M.\ Kachelrie\ss, I.\ V.\ Moskalenko,  and  S.\ S.\  Ostapchenko,
   Astrophys.\ J.\   {\bf 789},  136 (2014).

\bibitem{ko2012}
M.\ Kachelrie{\ss}  and  S.\ Ostapchenko,
Phys.\ Rev.\  D {\bf 86}, 043004 (2012).

\bibitem{ko2014}
M.\ Kachelrie{\ss}  and  S.\ Ostapchenko,
  Phys.\ Rev.\  D {\bf 90}, 083002  (2014).

\bibitem{kmo2015}
M.\ Kachelrie\ss, I.\ V.\ Moskalenko,  and  S.\ S.\  Ostapchenko,
   Astrophys.\ J.\   {\bf 803},  54 (2015);\\
I.\   Moskalenko, G.\ J\'ohannesson, M.\ Kachelrie\ss, E.\ Orlando, 
  S.\ S.\  Ostapchenko, and T.\ A.\ Porter,
POS ICRC2015  (2016) 495.

\bibitem{ost11}
S.\ Ostapchenko,   Phys.\ Rev.\  D {\bf 83}, 014018 (2011);
EPJ Web Conf.\  {\bf 52}, 02001 (2013).

\bibitem{boo1983}
C.\ N.\ Booth {\em et al.}, Phys.\ Rev.\ D {\bf 27}, 2018 (1983).

\bibitem{jae1973}
 K.\ Jaeger, J.\ Campbell, G.\ Charlton, D.\ Swanson, C.\ Fu,
H.\ A.\ Rubin, R.\ G.\ Glasser, D.\ Koetke, and J.\ Whitmore,
Phys.\ Rev.\ D {\bf 8}, 3824 (1973); {\em ibid.}  {\bf 11}, 1756 (1975).

\bibitem{lhcf-pi0}
O.~Adriani    {\em et al.}  (LHCf Collaboration),
 Phys.\ Rev.\  D  {\bf 94}, 032007 (2016).

\bibitem{na61} A.\ Aduszkiewicz  {\em et al.} (NA61/SHINE Collaboration),
   Eur.\ Phys.\ J.\  C \textbf{77}, 671 (2017).

   

\bibitem{aai2018} R.\ Aaij  {\em et al.} (LHCb Collaboration), 
   arXiv:1808.06127 [hep-ex].

\bibitem{gaisser}
For a text-book discussion, see T.~K.~Gaisser,
{\it Cosmic rays and particle physics}, 
(Cambridge University Press, Cambridge 1990).

\bibitem{blu2017}
K.~Blum, R.~Sato and M.~Takimoto,
  Phys.\ Rev.\ D {\bf 98}, no. 6, 063022 (2018).


\bibitem{den2011}
D.\ d'Enterria, R.\ Engel, T.\ Pierog, S.\ Ostapchenko, and K.\ Werner,
Astropart.\ Phys.\   {\bf 35}, 98 (2011).

\bibitem{obp2016}
 S.~Ostapchenko, M.\ Bleicher,  T.\ Pierog,  and K.~Werner, 
   Phys.\ Rev.\  D   {\bf 94},   114026 (2016).

\bibitem{dim2014}
M.\ di Mauro, F.\ Donato, A.\ Goudelis, and D.\ Serpico, 
  Phys.\ Rev.\  D {\bf 90}, 085017  (2014).

\end{thebibliography}
\end{document}